\documentstyle[12pt]{article}
\begin{document}
\newcommand{\be}{\begin{equation}}
\newcommand{\ee}{\end{equation}}
\newcommand{\bea}{\begin{eqnarray}}
\newcommand{\eea}{\end{eqnarray}}
\def\pd{\partial}
\def\no{\nonumber}
\baselineskip=24pt plus 2pt

\vspace{2mm}
\begin{center}
{\large \bf Stochastic Response, Cascading and Control of Colored Noise in Dynamical Systems } \\
\vspace{3mm}
Rue-Ron Hsu\footnote{E-mail address: rrhsu@mail.ncku.edu.tw},
Jyh-Long Chern\footnote{E-mail address: jlchern@mail.ncku.edu.tw},
Wei-Fu Lin\footnote{E-mail address: wfl@ibm65.phys.ncku.edu.tw},
and Chia-Chu Chen\footnote{E-mail address: chiachu@mail.ncku.edu.tw}\\
Nonlinear Science Group, Department of Physics,\\
National Cheng Kung University,\\ 
Tainan, Taiwan 70101, Republic of China\\
\end{center}
\vspace{2mm}

\begin{center}
{\bf ABSTRACT}
\end{center} 
~~~~We analytically determine the correlation functions of the stochastic 
response of a generic mapping system driven by colored noise. We also address the issue of noise 
cascading in coupled-element systems, particularly in a uni-directionally coupled system with a dc-filtered coupling.  
A noise-feedback method and an opposite-pairing method are
introduced to control the stochastic response. Excellent
agreements are obtained between analytical results and numerical simulations.\\
PACS number(s): 05.45.+b, 05.40.+j, 02.50.Fz.\\

\hfill{Typeset using \LaTeX}
\newpage 
\section{Introduction}

~~~~The stochastic response of dynamical systems to noise has attracted much
attention recently and is a recurrent theme in the study of nonlinear
systems. Some classic works were accomplished on the persistence of dynamical system under random perturbation~\cite{1}.  Recently, there are many novel and intriguing phenomena induced by noise,
for instance, noise-induced order~\cite{2} , synchronization by external 
noise~\cite{3} , stochastic resonance~\cite{4} , noise-driven coherence
resonance~\cite{5} , and noise-assisted stabilization~\cite{6}. Some experimental verifications of these phenomena have been found in extremely complicated real 
systems, such as the brain~\cite{7} and plasma~\cite{8}, and the physical picture can be captured by dynamical models with simple additive noise. On the other hand, noise-induced multistability has been addressed
in exploring the influence of multiplicative and additive noise~\cite{9},
and the reentrance phenomenon was introduced in studying the dynamical 
system with colored noise~\cite{10}.  The colored noise is the noise source with finite correlation time. In opposite to colored noise, white noise have not any correlation between time lapse.  Whenever the noise is 
dynamical involved, the correlation functions of the stochastic response 
of dynamical system to colored noise is usually very difficult to obtain exactly.
Several theories, for examples, the small correlation time
theory~\cite{11}, the functional-calculus theory~\cite{12}, the decoupling 
theory~\cite{13}, the unified colored-noise theory~\cite{14} and the 
interpolation procedure~\cite{15}, were proposed to treat the colored noise 
problem in stochastic processes. However,  analytical results can only 
be obtained under the limit of small or infinite correlation time. Therefore, how to determine the stochastic 
response of the colored noise exactly, is still an important issue. 

In another front, it is known that collective behaviour 
which arises from the coupling among the elements could be dramatically different 
from the behaviour of individual element, and is also a current interest in nonlinear dynamics.  
In the {\sl deterministic} systems, some novel features, such as self-induced spatial disorder and chaotic itinerancy, have been 
numerically found in distributed nonlinear element systems~\cite{16}. Universal critical behaviours have also been found in coupled map lattices~\cite{17} and the first order phase transition was studied in coupled oscillator systems~\cite{18}.  Nevertheless,
toward a more realistic model, one should include the noisy elements. Indeed,
 the stochastic resonance has been explored by numerical simulation and experiment in
 coupled-element systems with additive noise
~\cite{19}.  However, it is of fundamental interest to provide more illustrative example to clarify the exact role played by noise in the coupled system. 

In this paper, we analytically determine the correlation functions of the stochastic response of the steady state when the colored noise is dynamically involved in a generic mapping system. We also investigate the issue of noise cascading in coupled dynamical systems. As an analytical illustration, we adopt a special model in which all the elements are uni-directionlly
coupled to each other, specifically, we consider a dc-filtered coupling. The coupling is fundamental different to the global or nearest-neighbour coupling in spatially extended system[16,17], which is receiving much attention nowadays. But, this simple model thus serves as an analytically solvable model to illustrate the nature of noise during propagation in spatial extent under the influence of dynamical structures.  Furthermore, we found two 
elegant ways to control the stochastic response. 

\section{Stochastic response of colored noise}

~~~~To begin with, we consider
a generic one-dimension nonlinear map in which the noise is involved
dynamically,
\be 
x_{n+1}=f_r(x_n)+\zeta_n.
\ee
where $\zeta_n$ can be colored or white noise with zero mean fluctuation. For simplicity, we only treat 
the stochastic response of the fixed point,  one can extend
our results to higher periods by following the similar procedure in Reference~\cite{20}.
The dispersive output can also be expressed as 
\be
x_n=\overline{x}+\xi_n,
\ee
where $\overline{x}$ is the noise-free fixed point, $\overline{x}=
f_r(\overline{x})$, and $\xi_n$ is 
 the stochastic response.
From now, the amplitude of the noise and stochastic response are assumed to 
be small in comparison with the fixed point, such that the 
perturbation scheme is applicable. By substituting Eq.(2) into Eq.(1), a recurrence relation is obtained:
\be
\xi_{n+1}=M\xi_n+\zeta_n,
\ee
where $M=\frac{\partial f_r(x)}{\partial x}\vert_{\overline{x}}$.  
With $\left\vert M \right\vert <1$, we will , after the transient, have a small response fluctuation around the stable noise-free fixed point. By iterating Eq.(3), we have 
\be
\xi_k=M^k \xi_0 + \sum_{j=1}^k (M)^{j-1}\zeta_{k-j} 
\ee
After passing the transient state, i.e. $k>>1$, 
 the first term of Eq.(4) can be neglected. As a result,
the mean fluctuation of stochastic response is zero, if 
$<\zeta_k>=0$. Here, $<A_k>\equiv{\lim_{N\rightarrow\infty}}\frac{1}{N}
{\sum_{i=0}^{N-1}A_{k+i}}$ denotes the long time average of $A_{k+i}$. The index $k$ just reminds us that we are taking average for those states after the transient $k>>1$, and it is irrelevant to the average.  The zero mean of stochastic response imply that the average of dispersive output $x_n$ is same as the fixed point $\bar x$.
The general relations 
between the correlations of stochastic response and those of input noise can be obtained in compact forms:
\bea
<\xi_k \xi_k>=\frac{1}{1-M^2}\left \{ <\zeta_k \zeta_k >+2\sum_{i=1}^\infty M^i
<\zeta_k\zeta_{k+i}>\right \}\\
<\xi_k\xi_{k+\ell}>=\frac{M^\ell}{1-M^2}\left \{ <\zeta_k \zeta_k>+2\sum_{i=1}^\infty 
M^i <\zeta_k\zeta_{k+i}> \right \}\no\\
+\sum_{j=1}^\ell\sum_{i=0}^\infty 
M^{\ell-j+i}<\zeta_k\zeta_{k+j+i}> ,~~~~ \ell\geq 1.
\eea
If the correlations of the input noise are given, one can obtain 
the correlation of stochastic response after the summations are accomplished.
For example, if the input noise $\zeta_k$ is a white noise and is denoted by $\eta_k$ 
which have zero mean and derivation $D=<\eta^2>$, then the mean 
fluctuation of stochastic response is zero and the 
correlation function of the stochastic response is
\be
<\xi_k\xi_{k+\ell}>=\frac{M^\ell D}{1-M^2},~~~ \ell\geq 0.
\ee
Eq.(7) shows that the stochastic responses are colored when the white 
noise is dynamically convoluted by a generic map unless the system 
is critical i.e. $M=0$. The correlation time $\tau_c$ of the colored response can be identified as 
$\tau_c=-(\left\vert \ln{\left\vert M \right\vert } \right\vert )^{-1}$.

Next, we use the result of Eq.(7) as the input colored noise to drive another system,
$\tilde x_{n+1}=\tilde f_{r}(\tilde x_n)+\xi_n.$ 
With Eqs.(5) and (6), the mean fluctuation and the correlation 
of the stochastic responses, $\tilde \xi_n=\tilde x_n-\overline{\tilde x}$, of the new system could be deduced. They are $<\tilde \xi_k> = 0$, and 
\bea
<\tilde \xi_k \tilde \xi_k>&=&\frac {D} {(1-M^2)}  \left[ \frac{1}{(1-\tilde M^{ 2})}
\frac{(1+M \tilde M)}{(1-M\tilde M)} \right] ,\\
<\tilde \xi_{k} \tilde \xi_{k+\ell}>&=&\frac{D}{(1-M^2)} \biggl\lbrace
 \frac{1}{(1-M \tilde M)}\no\\ & &\times
\left[ M(\frac{\tilde M^{\ell}-M^\ell}{\tilde M -M})
 + \tilde M^{\ell}
(\frac{1+M \tilde M}{1-\tilde M^{2}}) \right] \biggr\rbrace,~~\ell\geq 1,
\eea
where $\vert\tilde M\vert=\vert \frac{\partial \tilde f_r}{\partial \tilde x}
\vert_{\overline{\tilde x}}\vert <1$ and 
$\overline {\tilde x}=\tilde f_{r}
(\overline {\tilde x})$.
The response and the correlation time are related to the dynamical structures $M$ and $\tilde M$ explicitly. Eqs.(8) and (9) also suggest that the correlation of the response could be greatly enhanced by the variation of dynamical structure. Referring to Fig.1, when the first stage is at the critical point ($M=0$), the stochastic response of the second stage exhibits a symmetry at the critical point ($\tilde M=0$). The symmetry is broken as the dynamics of the first stage moves away from its critical point. The minimum of stochastic response will shift to $M\tilde M<0$ region.  This fact will provide us a way to control the stochastic response and we will return to this point later.

Moreover, The extention of this result to higher periods case straightforward but tedious.  For example, to explore the stochastic response of each state of period-2, we shall divide the out data into two sets, called even and odd parts, and estimate the correlation for each set individually.  Then, the correlations of stochastic response on period-2 states can be simply carried out by converting $M$ and $\zeta_n$ to $M_2=\frac {\partial f}{\partial x}\vert_{\bar x_1}\frac{\partial f}{\partial x}\vert_{\bar x_2}$ and $b_n=\frac{\partial f}{\partial x}\vert_{\bar x'}\zeta_n + \zeta_{n+1}$ in Eqs.(5) and (6) respectively. Where $\bar x_1$ ( $\bar x_2$) is the mean value of odd (even) part of dispersive output, i.e. the period-2 orbit for noise-free case, and $\bar x'$ is $\bar x_2$ ( $\bar x_1$) when the index $n$ of $b_n$ is even (odd).

\section{Noise cascading}

~~~~As mentioned above, the noise cascading problem in coupled-element systems, especially for spatially extend coupled system, is remained as a challenge issue to be addressed. 
To gain more illustrative features, we take a dynamical system, called the first stage, which is driven 
by a white noise. The stochastic response is extracted by a dc-filter
and used as the noise to 
drive another system, called the second stage. This procedure could be repeated successively.  Physically, all the dynamics are described by
the following mapping, 
\bea
x_{n+1}^{(1)} & = & f_r^{(1)}(x_{n}^{(1)}) + \eta_n\no\\
x_{n+1}^{(2)} & = & f_r^{(2)}(x_{n}^{(2)}) + \xi_n^{(1)}\\
\ldots \no\\
x_{n+1}^{(m)} & = & f_r^{(m)}(x_{n}^{(m)}) + \xi_n^{(m-1)},\no
\eea
where $\xi_n^{(i)}=x_n^{(i)}-\overline x ^{(i)}$,
 and can be extracted by a dc-filter.  The essential feature of noise 
cascading can be addressed by treating identical systems (all $f^{(i)}_r$ are the same). To obtain 
an explicit form of analytical result we will assume 
$\vert M\vert <<1$.
Using Eqs. (5) and (6),  and keeping the lowest order of $M$, 
it can be proven that the correlation functions for the $m$-th $(m\geq 2)$ stage are 
\bea
<\xi_k^{(m)} \xi_k^{(m)}>&\approx&  (1+m^2 M^2) D \\
<\xi_k^{(m)} \xi_{k+\ell}^{(m)}>&\approx& C _{m-1} ^{\ell+m-1} M^\ell D 
,~~\ell\geq 1,
\eea
where $C _{m-1} ^{\ell+m-1}$ is the binomial coefficient.
The proof is shown in appendix.
These results indicate that the standard deviation of 
the noise is amplified when more stages are
included. As shown in Fig.2, the analytical result is also supported by numerical simulation .
We note that the correlation functions monotonously decrease with $\ell$ for $M>0$ and oscillating
decay for $M<0$. Both damping factors have $\tau_c=-(\vert \ln{\vert M \vert
}\vert )^{-1}$ for $\ell >>1$. 

\section{Control of noise response}
The noise amplification found above eventually leads to the breakdown of the coupled-element system as long as there are too many elements .  This amplification phenomenon  raises an important issue, namely, {\sl whether or not a controllable stochastic response can be obtained in noise cascading.} If the answer is negative then it is hopeless to study noise cascading in dynamical systems. Fortunately, we found two way to surmount the above difficulty.
The first solution to this problem can be achieved by means of feedback.
Let us modify the
generic one-dimension map, Eq.(1), by adding feedback response $a\xi_n$,
\be
y_{n+1}=f_r(y_n)+\zeta_n+a \xi_n,
\ee 
where $a$ is the feedback strength and $\xi_n=y_n-\bar y$.
By using the same approach as before, we find the 
modified general relation between the correlations of stochastic 
response and those of input noise,
\bea
<\xi_k \xi_k>=\frac{1}{1-(M+a)^2} \left \{ <\zeta_k^2> + 2 \sum_{i=1}^{\infty}
(M+a)^i <\zeta_k\zeta_{k+i}> \right \}\\
<\xi_k \xi_{k+\ell}>=\frac{(M+a)^\ell}{1-(M+a)^2} \left \{ <\zeta_k^2> +
2 \sum_{i=1}^{\infty}(M+a)^i <\zeta_k\zeta_{k+i}> \right \} \no\\
+ \sum_{j=1}^{\ell} \sum^{\infty}_{i=0} (M+a)^{\ell-j+i} <\zeta_k\zeta_{k+j+i}>
, ~~ \ell \geq 1.
\eea
This result indicates that the output response at best could be reduced to the input noise if we adjust $a$ to be $a=-M$. Fig.3 is the simulation result and 
shows that the feedback scheme can reduce the stochastic response efficiently. The standard derivation of output response is minimized at $a=-M$, and it is indeed smaller than the case without feedback ($a=0$). On the other hand, as implied by the results of Eqs.(8) and (9), some suitable variations of dynamical structures may be employed to reduce 
the stochastic response, and thus postpone the breakdown. For a  simple illustration, let us consider a two-stage system with opposite pairs. It means that we choose the stability quantity of second stage to be $\tilde M=-M$, which is opposite to the previous one.  By using Eq.(8) and (9), the correlation of the output noise is 
\be
<\tilde\xi_k \tilde \xi_{k+2\ell}>=\frac{M^{2\ell}D}{1-M^4} ,~~~ \ell\geq 0,
\ee
and $<\tilde\xi_k \tilde \xi_{k+2\ell +1}>=0$.  Since $\vert M\vert <1$, 
it is clear that the output noise is suppressed in comparison with  Eq.(7) and the correlation time is still $\tau_c=-(\left\vert \ln{ \vert M\vert } \right\vert )^{-1}$ in this case.  That offers us an opportunity to reduce the colored noise by choosing the dynamical structure with opposite pairs.  Extending this opposite-pairing method to the cascading model with $p$-pair, $M^{(2p-1)}=M$ and $M^{(2p)}=-M$, $p\geq 1$, one finds that the correlation functions for the $p$-th pair are 
\bea
<\xi_k^{(2p-1)} \xi_k^{(2p-1)}>&\approx&  (1+M^2+p^2 M^4) D\\
<\xi_k^{(2p)} \xi_k^{(2p)}>&\approx& (1+p^2 M^4) D .
\eea
These results can be proven by following the same procedure shown in the appendix.
 As shown in Fig.4, the output noise is indeed greatly reduced and the breakdown is postponed. 

\section{Concluding remarks}

~~~~In summary, we analytically derive the relation between the correlation of stochastic response of a steady state in a mapping system and that of input noise.  We also obtain the exact form of the correlations of the noise cascading in an uni-directionally dc-filtered coupled system.  Our results suggest that noise amplification might lead to catastrophic disaster in coupled-element systems. However, in certain cases, such as the one discussed above, the amplification   can be controlled either by a feedback scheme or by an opposite-pairing arrangement of dynamical structures. Let us address the implication on this point.  Assume that the noise-cascading system will breakdown when the derivation of stochastic response is large than a threshold $D_{th}$, thus the noise amplification set the limit on the cascading length. For examples, the maximum lengths of the identical-coupled and opposite-pairing cascading models are 
$m_{max}=[ \sqrt{\frac {(D_{th}-D)} {D}} \frac {1}{\vert M\vert} ],$ and
$2p_{max}= 2[ \sqrt{\frac {(D_{th}-D)} {D}} \frac {1}{M^2}] $ respectively, where $[~~]$ is the Gaussian notation.
Since $\vert M \vert<1$, we will have a shorter chain for identical elements and a longer chain for the opposite-pairing one. On the other hand, the noise-feedback coupling will give an un-limited chain. Thus, it is worthwhile noting that our models, even they are not so realistic now, may provide a scheme in which the noise could influence the size of the pattern formation~\cite{21}.

Finally, there are some remarks should be mentioned.
All the analytical results agree with the numerical simulations very well as long as the ratio between the amplitude
of input noise and the value of fixed point is
less than 5\%, and the dynamical systems are not
too close to the bifurcation point. The same conclusion can also be drawn for the Ikeda map. 
It is important to note that the dc-filtered coupling is essential through out this investigation. Without the filter, the dynamical structures of each stage will be greatly altered by the dc-signal and the coupled system will quickly breakdown. 

\vspace{20mm}
\begin{center}
{\large \bf Acknowledgements}
\end{center}
\vspace{10mm}
We thank Prof. H.-T. Su for usefull discussions and reading the manuscript. 
The work is partially supported by the National Science Council,
Taiwan, R.O.C. under the contract numbers. NSC 86-2112-M006-003 and NSC 86-2112-M006-008.

\newpage
\begin{center}
{\large \bf Appendix}
\end{center}
\vspace{5mm}

Here, we will use mathematical induction to prove Eqs.(11) and (12).  After appling the approximation $\vert M\vert <<1$  and keeping those leading terms,  Eqs. (5)-(6) end up with two general relations between the $(m+1)$-th stage correlation functions and $m$-th stage correlation functions, i.e.,
\bea
<\xi_k^{(m)} \xi_k^{(m)}>&\approx&  (1+ M^2) <\xi_k^{(m-1)} \xi_k^{(m-1)}> +2M<\xi_k^{(m-1)} \xi_{k+1}^{(m-1)}>,~~~~~~\\
<\xi_k^{(m)} \xi_{k+\ell}^{(m)}>&\approx& M^\ell <\xi_k^{(m-1)} \xi_k^{(m-1)}>\no\\ 
&~~~& ~+ \sum_{i=1}^{\ell} M^{\ell-i} <\xi_k^{(m-1)} \xi_{k+i}^{(m-1)}> 
,~~~\ell\geq 1.~~
\eea
We note that the leading order of $<\xi_k^{(m-1)} \xi_{k+i}^{(m-1)}>$ is $M^i$. 

Let us first consider the $m=2$ case.  We read out $<\xi_k^{(1)} \xi_k^{(1)}>\approx  (1+ M^2) M^\ell D$ from Eq.(7), and put it into Eqs.(19) and (20). It is easy to check that
\bea
<\xi_k^{(2)} \xi_k^{(2)}>&\approx&  (1+4 M^2) D \\
<\xi_k^{(2)} \xi_{k+\ell}^{(2)}>&\approx& {\ell+1} M^\ell D=C _{1} ^{\ell+1} M^\ell D .
,~~\ell\geq 1.
\eea
Next, for the cases of $m>2$, we will show that if Eqs. (11) and (12) are true for $m$-th stage then they are also true for the $m+1$ case.   Plugging Eqs. (11)-(12) into Eqs. (19)-(20) and expanding them to those leading order terms, we obtain
\bea
<\xi_k^{(m+1)} \xi_k^{(m+1)}>&\approx& (1+M^2) (1+m^2 M^2) D 
+2 C _{m-1} ^{m} M^2 D \no\\  
&\approx& (1+(m+1)^2 M^2) D \\ 
<\xi_k^{(m+1)} \xi_{k+\ell}^{(m+1)}>
&\approx& M^\ell (1+m^2 M^2) D+\sum_{i=1}^{\ell} C _{m-1} ^{i+m-1} M^\ell D \no\\ 
&\approx& \sum_{i=0}^{\ell} C _{m-1} ^{i+m-1} M^\ell D 
= C _{(m+1)-1} ^{\ell+(m+1)-1} M^\ell D
,~~\ell\geq 1,
\eea
where the summation identity of binomial coefficients is used.
By mathematical induction, Eqs. (11) and (12) are true for any $m$-th stage.  Q.E.D.

\newpage
\begin{flushleft}
{\Large\bf Figure Captions :}\\
\end{flushleft}
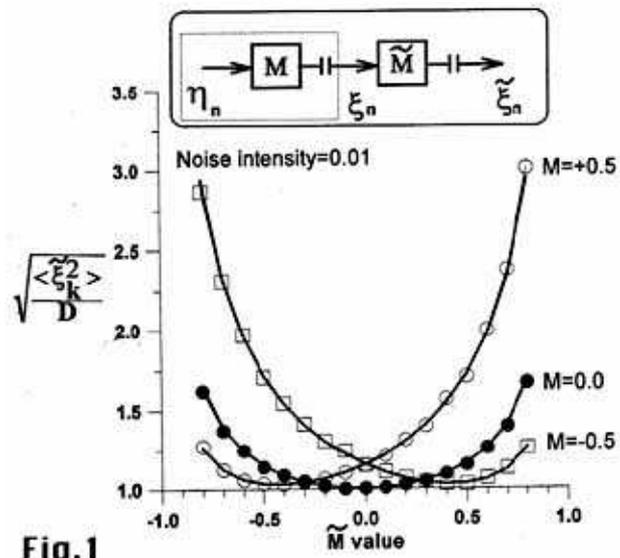
{\bf Figure 1.}\\
{\bf Up:} The block diagram for the stochastic response $\tilde \xi_n$ of the dynamical system $\tilde M$ where the correlation function of the input noise is Eq.(7). That is equivalent to a uni-directionally coupled system in which two elements are coupled by a dc-filter and with white noise input. \\
{\bf Bottom:} $\sqrt {\frac {<\tilde \xi_k^2>}{D}}$ via $\tilde M$ for several fixed $M$. The solid lines denote the analytical estimations. The circle (bullet, box) denote the simulation results for the dynamical structures $M=0.5 ~(0.0, -0.5)$ of first stage. Here, the dynamical systems are logistic maps $f_r (x)=rx(1-x)$ with $M=2-r$, and $\tilde f_{\tilde r}(\tilde x)=\tilde r\tilde x(1-\tilde x)$ with $\tilde M=2-\tilde r$.  The input uniform white noise $\eta_n$ have amplitude 0.01.\\
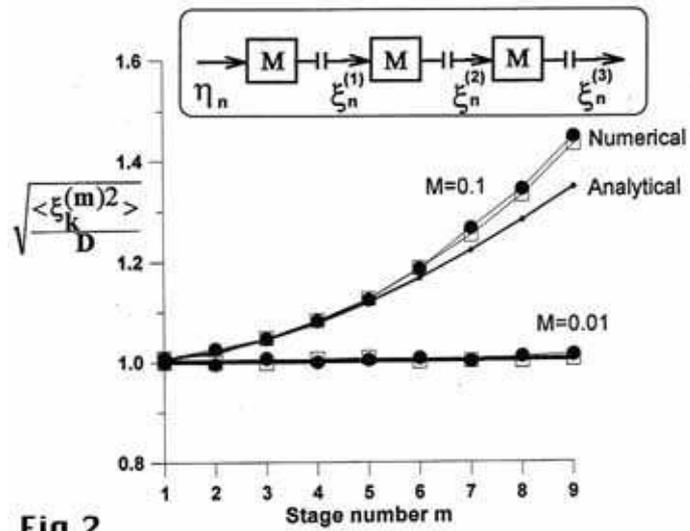
{\bf Figure 2.}  \\
{\bf Up:} The block diagram for an identical coupled-element system. \\
{\bf Bottom:} $\sqrt {\frac {<\xi_k^{(m)2}>}{D}}$ via stage number $m$, for two identical coupled-element systems which have different elements $f_r (x)=1.9*x(1-x)$ and $f_r (x)=1.99*x(1-x)$, respectively. The solid lines denote the analytical results Eq.(11). The bullet ( box) denote the simulation results for the input noises with amplitudes 0.01 (0.001).\\
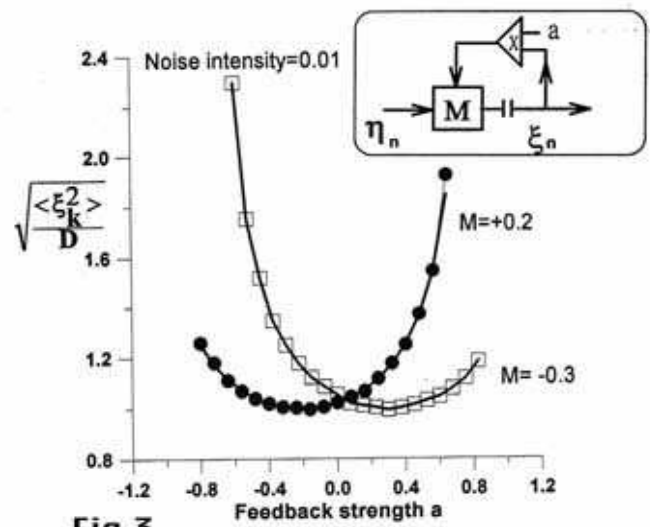
{\bf Figure 3.}  \\
{\bf Up:} The block diagram for a feedback scheme to reduce the stochastic response. \\
{\bf Bottom:} $\sqrt {\frac {<\xi_k^{2}>}{D}}$ via the feedback strength $a$.  The input uniform white noise has  amplitude 0.01. Two logistic maps $f_r (x)=1.8*x(1-x)$ with $M=0.2$ and $f_r (x)=2.3*x(1-x)$ with $M=-0.3$ are demonstrated.  Both simulations and analytical results show that the responses are minimized at $a=-0.2$ and $a=0.3$ respectively.\\
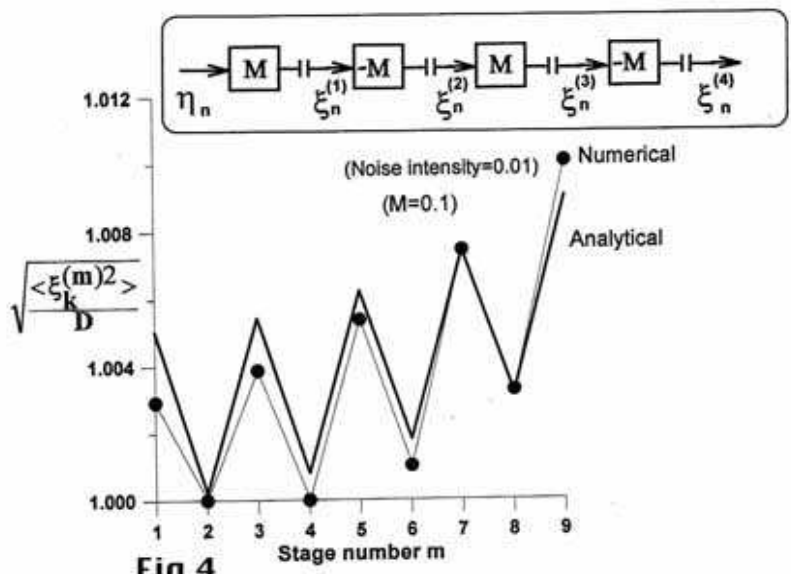
{\bf Figure 4.}  \\
{\bf Up:} The block diagram for a opposite-pairing scheme to reduce the stochastic response.  \\
{\bf Bottom:} $\sqrt {\frac {<\xi_k^{(m)2}>}{D}}$ via stage number $m$, for the opposite-pairing coupled system which have a pair of logistic maps $f_r (x)=1.9*x(1-x)$ and $f_r (x)=2.1*x(1-x)$. The solid lines denote the analytical results Eqs.(17) and (18). The bullet denote the simulation result for the input noise with amplitudes 0.01 .\\
\end{document}